\documentclass[amsmath,pra,preprint,showpacs]{revtex4}
\usepackage{graphics}

\begin{document} 

\title{Effect of an atom on a quantum guided field in a weakly driven fiber-Bragg-grating cavity}

\author{Fam Le Kien}
\altaffiliation{Also at Institute of Physics, Vietnamese Academy of Science and Technology, Hanoi, Vietnam.}
\affiliation{Department of Applied Physics and Chemistry, 
University of Electro-Communications, Chofu, Tokyo 182-8585, Japan}

\author{K. Hakuta} 
\affiliation{Department of Applied Physics and Chemistry, 
University of Electro-Communications, Chofu, Tokyo 182-8585, Japan}

\date{\today}

\begin{abstract}
We study the interaction of an atom with a quantum guided field in a weakly driven fiber-Bragg-grating (FBG) cavity. We present an effective Hamiltonian and derive the density-matrix equations for the combined atom-cavity system. We calculate the mean photon number, the second-order photon correlation function, and the atomic excited-state population. We show that, due to the confinement of the guided cavity field in the fiber cross-section plane and in the space between the FBG mirrors, the presence of the atom in the FBG cavity can significantly affect the mean photon number and the photon statistics 
even though the cavity finesse is moderate, the cavity is long, and the probe field is weak.   
\end{abstract}

\maketitle

\section{Introduction}

In cavity quantum electrodynamics, an interesting regime, called the regime
of strong coupling, occurs when the maximal atom-field dipole coupling strength exceeds the cavity field decay rate and the atomic spontaneous emission rate \cite{Berman}. In this regime, excitations can be exchanged coherently between the atom and the field several times before the incoherent decay process occurs, the properties of the field can be significantly modified by the presence of even a single atom, and the presence of a single photon in the field can saturate the response of the atom.
The effects of single atoms on the cavity field in real time \cite{Rempe,Doherty} have been observed \cite{Mabuchi,Hood,Munstermann}. It has been reported that the presence of an atom in the cavity, which is tuned to the atomic transition and resonantly driven by a laser field, can lead to a dramatic drop in the transmitted intensity \cite{Mabuchi}. It has been demonstrated  \cite{Hood Science, Pinkse Nature} that the spatial variation of the cavity mode can lead to a confining potential sufficient to trap an atom within the cavity mode even for a single quantum of excitation \cite{Haroche, Englert,Doherty 2000}. Cooling \cite{Ritsch} of single atoms with single photons in a high-$Q$ cavity has also been investigated. 

Recently, it has been proposed to combine the cavity technique with the nanofiber technique to obtain a hybrid system, where the interaction is enhanced by the transverse confinement of the field in the fiber cross-section plane as well as the longitudinal confinement of the field between the mirrors. 
It has been shown that the presence of a fiber-Bragg-grating (FBG) cavity with a large length (on the order of 10 cm) and a moderate finesse (about 30) can significantly enhance the group delay of a guided probe field \cite{fibercavity} and substantially enhance the channeling of emission from an atom into a nanostructure \cite{cavityspon}. There has been a large body of work involving fiber Bragg gratings over the past two decades \cite{Othenos,Kashyap,Canning,Wan,Chow,Gupta}. With careful control of the grating writing process and appropriate choice of glass material, a FBG resonator can have a finesse of well over 1000 and a linewidth of a few MHz \cite{Gupta}. It is worth mentioning that several methods for trapping and guiding neutral atoms outside a fiber have been proposed and studied 
\cite{Dowling,onecolor,twocolors,Lee,Rauschenbeutel,focus trap,twocolor experiment}. 
A trapping method based on the use of two (red- and blue-detuned) light beams has been studied for large-radius fibers \cite{Dowling} and nanofibers \cite{twocolors} and has recently been experimentally realized \cite{twocolor experiment}.

In this paper, we study the interaction of an atom with a quantum guided field in a weakly driven FBG cavity. We show that, due to the confinement of the guided cavity field in the fiber cross-section plane and in the space between the FBG mirrors, the presence of the atom in the FBG cavity can significantly affect the mean photon number as well as the photon statistics even though the cavity finesse is moderate, the cavity is long, and the probe field is weak. 

The paper is organized as follows. In Sec.\ \ref{sec:model} we describe the model. In Sec.\ \ref{sec:equation} we derive the density-matrix equations for the combined atom-field system. 
In Sec.\ \ref{sec:results} we  present the results of numerical calculations for the mean photon number, the second-order correlation function, and the atomic excited-state population. Our conclusions are given in Sec.~\ref{sec:summary}.

\section{Model}
\label{sec:model}

\begin{figure}[tbh]
\begin{center}
 \includegraphics{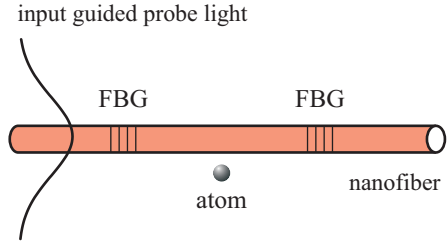}
 \end{center}
\caption{(Color online) An atom in the vicinity of a nanofiber with two 
fiber-Bragg-grating mirrors driven by a weak guided probe light field.}
\label{fig1}
\end{figure}

We consider a two-level atom in the vicinity of a nanofiber with two FBG mirrors (see Fig. \ref{fig1}). The field in the guided modes of the nanofiber is reflected back and forth between the FBG mirrors. The nanofiber has a cylindrical silica core of radius $a$ and of refractive index $n_1=1.45$ and an infinite vacuum clad of refractive index $n_2=1$. In view of the very low losses of silica in the wavelength range of interest, we neglect material absorption. We use the cylindrical coordinates $(r,\varphi,z)$, with $z$ being the axis of the fiber. 

In the presence of the fiber, the electromagnetic field can be decomposed into guided and radiation modes \cite{fiber books}. In order to describe the field in a quantum mechanical formalism, we follow the continuum field quantization procedures presented in \cite{Loudon}. First, we temporally neglect the presence of the FBG mirrors. Regarding the guided modes, we assume that the single-mode condition \cite{fiber books} is satisfied for a finite bandwidth around the atomic transition frequency $\omega_0$. We label each fundamental guided mode HE$_{11}$ with a frequency $\omega$ in this bandwidth by an index $\mu=(\omega,f,l)$, where $f=+,-$ denotes the forward or backward propagation direction
and $l=+,-$ denotes the counterclockwise or clockwise rotation of polarization. The quantum expression for the electric positive-frequency component $\mathbf{E}^{(+)}_{\mathrm{gyd}}$ of the field in the fiber guided modes is \cite{cesium decay}
\begin{equation}\label{1}
\mathbf{E}^{(+)}_{\mathrm{gyd}}=i\sum_{\mu}\sqrt{\frac{\hbar\omega\beta'}{4\pi\epsilon_0}}
\;a_{\mu}\mathbf{e}^{(\mu)}e^{i(f\beta z+l\varphi)-i\omega t}.
\end{equation}
Here $\mathbf{e}^{(\mu)}=\mathbf{e}^{(\mu)}(r,\varphi)$ is the profile function of the guided mode $\mu$ in the classical problem, $a_{\mu}$ is the corresponding photon annihilation operator, 
$\sum_{\mu}=\sum_{fl}\int_0^{\infty}d\omega$ is the summation over the guided modes,
$\beta$ is the longitudinal propagation constant, and $\beta'$ is the derivative of $\beta$
with respect to $\omega$. The constant $\beta$ is determined by the
fiber eigenvalue equation \cite{fiber books}. The operators $a_{\mu}$ and $a_{\mu}^\dagger$ satisfy the continuous-mode bosonic commutation rules $[a_{\mu},a_{\mu'}^\dagger]=\delta(\omega-\omega')\delta_{ff'}\delta_{ll'}$. The explicit expression for the mode function $\mathbf{e}^{(\mu)}$ is given
in Refs. \cite{cesium decay,fiber books}. According to Ref. \cite{cesium decay}, the rate of spontaneous emission into guided modes is given by
\begin{equation}\label{13a}
\gamma_{\mathrm{gyd}}=\frac{\omega_0}{2\epsilon_0\hbar v_g}\sum_{fl}
\big|\mathbf{d}\cdot\mathbf{e}^{(\omega_0,f,l)}\big|^2.
\end{equation}
Here $\mathbf{d}$ is the matrix element of the electric dipole moment of the atom and $v_g=1/\beta'(\omega_0)$ is the group velocity of the guided field.

Regarding the radiation modes, the longitudinal propagation constant $\beta$ for each frequency $\omega$ can vary continuously, from $-k$ to $k$, with $k=\omega/c$ being the wave number. 
We label each radiation mode by an index $\nu=(\omega,\beta,m,l)$, where 
$m=0,\pm1,\pm2,\dots$ is the mode order and $l=+,-$ is the mode polarization. The quantum expression for the electric positive-frequency component 
$\mathbf{E}^{(+)}_{\mathrm{rad}}$ of the field in the radiation modes is \cite{cesium decay}
\begin{equation}
\mathbf{E}^{(+)}_{\mathrm{rad}}=i\sum_{\nu}
\sqrt{\frac{\hbar\omega}{4\pi\epsilon_0}}\;a_{\nu}\mathbf{e}^{(\nu)}e^{i(\beta z+m\varphi)-i\omega t}.
\label{4n}
\end{equation}
Here $\mathbf{e}^{(\nu)}=\mathbf{e}^{(\nu)}(r,\varphi)$ is the profile function of the radiation mode $\nu$ in the classical problem, $a_{\nu}$ is the corresponding photon annihilation 
operator, and $\sum_{\nu}=\sum_{ml}\int_0^{\infty}d\omega\int_{-k}^{k}d\beta$ is the summation over the radiation modes. The operators $a_{\nu}$ and $a_{\nu}^\dagger$ satisfy the continuous-mode bosonic commutation rules $[a_{\nu},a_{\nu'}^\dagger]=\delta(\omega-\omega')\delta(\beta-\beta')
\delta_{mm'}\delta_{ll'}$. The explicit expression for the mode function $\mathbf{e}^{(\nu)}$ is given
in  Refs. \cite{cesium decay,fiber books}. According to Ref. \cite{cesium decay}, the rate of spontaneous emission into radiation modes is given by
\begin{equation}\label{13b}
\gamma_{\mathrm{rad}}=\frac{\omega_0}{2\epsilon_0\hbar}\sum_{ml}\int _{-k_0}^{k_0}d\beta\,
\big|\mathbf{d}\cdot\mathbf{e}^{(\omega_0,\beta, m, l)}(r,\varphi)\big|^2.
\end{equation}

Next, we take into account the effect of the FBG mirrors on the mode functions. We assume that the two FBG mirrors are identical, having the same complex reflection and transmission coefficients $R$ and $T$, respectively, for the guided modes in a broad bandwidth around the atomic transition frequency $\omega_0$. In general, we have $|R|^2+|T|^2\leq 1$, where the equality (inequality) occurs for lossless (lossy) gratings. Without loss of essential physics, we assume that the gratings are lossless, that is, $|R|^2+|T|^2=1$. Let the mirrors be located at the positions $z=\pm L/2$ along the fiber, where $L$ is the distance between the mirrors. The guided modes are modified by the
presence of the mirrors. The mode functions of the cavity-modified guided modes are obtained, as usual in the Fabry-Perot theory, by summing the geometric series resulting from the multiple reflections by the mirrors \cite{Martini,Bjork,Cook}. Inside the cavity, the mode functions of the cavity-modified guided modes are given by
\begin{equation}\label{2}
\tilde{\mathbf{e}}^{(\omega,f,l)}=\mathbf{e}^{(\omega,f,l)}\frac{T}{1-R^2e^{2i\beta L}}
+\mathbf{e}^{(\omega,-f,l)}\frac{TR e^{i\beta(L-2fz)}}{1-R^2e^{2i\beta L}},
\end{equation}
and, hence, the electric positive-frequency component of the field in the cavity-modified guided modes is
\begin{equation}\label{3}
\mathbf{E}^{(+)}_{\mathrm{cavgyd}}=i\sum_{\mu}\sqrt{\frac{\hbar\omega\beta'}{4\pi\epsilon_0}}
\;a_{\mu}\tilde{\mathbf{e}}^{(\mu)}e^{i(f\beta z+l\varphi)-i\omega t}.
\end{equation}

We assume that the FBG mirrors do not reflect the radiation modes. This approximation is reasonable when the distance $L$ between the FBG mirrors is large as compared to the fiber radius $a$
and to the wavelength $\lambda_0=2\pi/k_0$, with $k_0=\omega_0/c$ being the wave number of the atomic transition. In the framework of this approximation, the mode functions of the radiation modes are not affected by the presence of the FBG mirrors. 

We drive the FBG cavity by a classical probe light field propagating along the fiber in a guided mode $\mu_p=(\omega_p, f_p, l_p)$. Let $P_{\mathrm{in}}$ be the incident power. The transmitted power is given by
\begin{equation}\label{4a}
P_{\mathrm{out}}=P_{\mathrm{in}}\frac{(1-|R|^2)^2}{(1-|R|^2)^2+4|R|^2\sin^2\Theta(\omega_p)}.
\end{equation}
Here $\Theta(\omega)=\beta(\omega) L+\phi_R$ is the phase shift caused by a single cavity crossing and a single reflection, with $\phi_R$ being the phase of the reflection coefficient $R$, that is, $R=|R|e^{i\phi_R}$. We assume that the probe field frequency $\omega_p$ is close to a resonant cavity frequency $\omega_c$, which is determined by the equation $\Theta(\omega_c)=m\pi$, with $m$ being an integer number. To the first order in $\omega_p-\omega_c$, we have the expansion $\Theta(\omega_p)=m\pi+(L/v_g)(\omega_p-\omega_c)$. Hence, we find
\begin{equation}\label{5a}
P_{\mathrm{out}}=P_{\mathrm{in}}\frac{(1-|R|^2)^2}{(1-|R|^2)^2+4|R|^2(L/v_g)^2(\omega_p-\omega_c)^2}.
\end{equation}

In the framework of the input-output formulation for optical cavities, the evolution of the photon operator $a$ for the field in an empty two-sided cavity is governed by the equation \cite{Walls}
\begin{equation}\label{6}
\dot{a}=-i\omega_c a-\frac{\kappa}{2} a+\sqrt{\kappa/2}\,a_{\mathrm{in}}
+\sqrt{\kappa/2}\,b_{\mathrm{in}}.
\end{equation}
Here $a_{\mathrm{in}}$ and $b_{\mathrm{in}}$ are the input photon operators for the left and right sides of the cavity, respectively, and $\kappa$ is the cavity damping coefficient, which is assumed to be the same for the two sides. 
When the input field $a_{\mathrm{in}}$ is an external coherent monochromatic probe field, with the frequency $\omega_p$ and the power $P_{\mathrm{in}}=\hbar\omega_p\langle a_{\mathrm{in}}^{\dagger}a_{\mathrm{in}}\rangle$, and the input field $b_{\mathrm{in}}$ is in the vacuum state, the mean number 
$\bar{n}=\langle a^\dagger a\rangle$ of photons in the cavity is given by \cite{Walls} 
\begin{equation}\label{10}
\bar{n}=\frac{\eta^2}{\kappa^2/4+\Delta_c^2}.
\end{equation}
Here $\Delta_c=\omega_p-\omega_c$ is the detuning of the probe field from the cavity resonance
and 
\begin{equation}\label{10a}
\eta=\sqrt{\frac{\kappa}{2} \frac{P_{\mathrm{in}}}{\hbar\omega_p}}
\end{equation}
is the cavity pumping rate. 

It is clear that the interaction of the quantum cavity field with the external classical probe field can be described by the Hamiltonian
\begin{equation}\label{17}
H_P=-i\hbar(\eta e^{i\omega_pt} a-\mathrm{H.c.}).
\end{equation}
Note that the power of the transmitted field is related to the mean intracavity photon number $\bar{n}$ via the formula \cite{Walls}
\begin{equation}\label{12c}
P_{\mathrm{out}}=\frac{1}{2}\hbar\omega_p\kappa\bar{n}.
\end{equation}
When we insert Eq. (\ref{10}) into Eq. (\ref{12c}) and compare the result with Eq. (\ref{5a}), we find
\begin{equation}\label{11}
\kappa=\frac{(1-|R|^2)v_g}{|R|L}
\end{equation}
and
\begin{equation}\label{12e}
\eta=\sqrt{\frac{(1-|R|^2)v_gP_{\mathrm{in}}}{2|R|L\hbar\omega_p}}.
\end{equation}

The external probe field excites the guided cavity mode $\mu_c=(\omega_c,f_c,l_c)$,
where $f_c=f_p$ and $l_c=l_p$. In the single-mode regime,
the electric positive-frequency component of the field in the excited guided cavity mode is given by the expression
\begin{eqnarray}\label{14a}
\mathbf{E}^{(+)}_{\mathrm{cav}}&=&\frac{i}{2}\sqrt{\frac{\hbar\omega_c}{\epsilon_0 L}}
\;a\Big[\mathbf{e}^{(\omega_c,f_c,l_c)}e^{if_c\beta_c z}
\nonumber\\&&\mbox{}
+e^{im\pi}\mathbf{e}^{(\omega_c,-f_c,l_c)}e^{-if_c\beta_c z}\Big]e^{il_c\varphi}e^{-i\omega_c t}.
\end{eqnarray}
Hence we find that the interaction between the atom and the quantum guided cavity field in the dipole and rotating-wave approximations is described by the Hamiltonian 
\begin{equation}\label{15}
H_{AF}=-i\hbar (G a \sigma^\dagger-\mathrm{H.c.}),
\end{equation}
where $\sigma$ and $\sigma^{\dagger}$ are the downward and upward operators for the atomic transitions,
respectively, and 
\begin{eqnarray}\label{16}
G&=&\frac{1}{2}\sqrt{\frac{\omega_c}{\epsilon_0\hbar L}}
\mathbf{d}\cdot\Big[\mathbf{e}^{(\omega_c,f_c,l_c)}e^{if_c\beta_c z}
\nonumber\\&&\mbox{}
+e^{im\pi}\mathbf{e}^{(\omega_c,-f_c,l_c)}e^{-if_c\beta_c z}\Big]e^{il_c\varphi}
\end{eqnarray}
is the coupling coefficient.
Here $r$, $\varphi$, and $z$ are the cylindrical coordinates of the position of the atom.

In general, the atomic dipole vector $\mathbf{d}$ of a realistic atom is a complex vector.
We use the notation $V_0=V_z$ and $V_{\pm 1}=\mp(V_x\pm i V_y)/\sqrt{2}$ for the spherical components of an arbitrary vector $\mathbf{V}$. We assume that only one spherical component $d_q=d$ of the dipole vector $\mathbf{d}$, where $q=-1$, 0, or 1, is nonzero. Then, we have
$\mathbf{d}\cdot \mathbf{e}^{(\omega, f, l)}=(-1)^q d e_{-q}^{(\omega, f, l)}$.
Due to the properties of the mode profile functions \cite{fiber books}, we have
$e_{-q}^{(\omega, f, l)}(r,\varphi)=f^{1+q}e^{iq(\pi/2-\varphi)}|e_{-ql}^{(\omega)}(r,\varphi)|$.
Here we have introduced the notation $|e_0^{(\omega)}|=|e_z^{(\omega,+,+)}|$ and $|e_{\pm 1}^{(\omega)}|=(|e_r^{(\omega,+,+)}|\mp |e_{\varphi}^{(\omega,+,+)}|)/\sqrt{2}$. 
Furthermore, we assume that the probe field and hence the cavity field are counterclockwise circularly polarized, that is, $l_p=l_c=+$, and that the atomic transition is $\sigma^+$ polarized, that is, $q=1$.
Then, we have
$G=e^{-i(m+1)\pi/2}(\omega_cd^2/\epsilon_0\hbar L)^{1/2}
\,|e_{-1}^{(\omega_c)}|\,\cos{(\beta_c z+m\pi/2)}$.
We can remove the phase factor $e^{-i(m+1)\pi/2}$ by performing the transformation
$ae^{-i(m+1)\pi/2}=\tilde{a}$ for the photon operators or the transformation $\sigma e^{i(m+1)\pi/2}=\tilde{\sigma}$ for the atomic operators. Therefore, we can take the following expression
for the atom-field coupling coefficient:
\begin{eqnarray}\label{16f}
G=\sqrt{\frac{\omega_cd^2}{\epsilon_0\hbar L}}
\,|e_{-1}^{(\omega_c)}|\,\cos{(\beta_c z+m\pi/2)}.
\end{eqnarray}
Note that $\Omega=2G$ is sometimes called the vacuum Rabi frequency.

In the vicinity of the fiber surface, the atom experiences the effect of the van der Waals potential
on the internal state energy and on the center-of-mass motion. Let $V_g$ and $V_e$ be the
van der Waals potentials for the ground state $|g\rangle$ and the excited state $|e\rangle$, respectively. In the presence of the fiber, the atomic transition frequency is shifted from 
the bare frequency $\omega_0$ and is given by
\begin{equation}\label{32a}
\omega_a(r)=\omega_0+V_{eg}(r)/\hbar,
\end{equation}
where $V_{eg}=V_e-V_g$. The Hamiltonians of the atom and the guided cavity field in the case of no coupling between them are given by
\begin{equation}\label{30a}
H_A=\frac{\mathbf{p}^2}{2M}+\frac{1}{2}\hbar\omega_a\sigma_z+\frac{V_e+V_g}{2}
\end{equation}
and
\begin{equation}\label{30b}
H_F=\hbar\omega_c a^\dagger a,
\end{equation}
respectively. Here $\mathbf{p}$ and $M$ are the momentum and the mass of the atom, respectively.
When we sum up the Hamiltonians (\ref{17}), (\ref{15}), (\ref{30a}), and (\ref{30b}),
we obtain the total Hamiltonian $H=H_A+H_F+H_{AF}+H_P$ for the combined atom-field system. We decompose $H$ into two parts,
$H_0=\hbar\omega_p(a^\dagger a+\sigma_z/2)$ and $H_I=H-H_0$. In the interaction picture,
the combined atom-field system is described by the 
Hamiltonian $H_{\mathrm{int}}=e^{iH_0t/\hbar}H_Ie^{-iH_0t/\hbar}$, whose explicit expression is
\begin{eqnarray}\label{30}
H_{\mathrm{int}}&=&\frac{\mathbf{p}^2}{2M} -\frac{1}{2}\hbar\Delta_a\sigma_z-\hbar\Delta_c a^\dagger a
-i\hbar G(a \sigma^\dagger-a^\dagger \sigma)
\nonumber\\&&\mbox{} 
-i\hbar\eta( a-a^{\dagger})+\frac{V_e+V_g}{2}.
\end{eqnarray}
Here $\Delta_a=\omega_p-\omega_a$ is the detuning of the probe field frequency $\omega_p$ from the surface-shifted atomic transition frequency $\omega_a$. The Hamiltonian (\ref{30}) is almost same as that for a weakly driven microcavity \cite{Ritsch}. The difference between the two models is that the presence of the van der Waals potential in the case of the nanofiber-based cavity affects the transition frequency and the center-of-mass motion of the atom.

For the treatments in the remaining part of this paper, we assume that the translational motion of the atom can be neglected. This situation can be realized in the cases where the atom is trapped, the atom is slow (the atomic ensemble is kept at a low temperature), or the two-level atom is replaced by a heavy particle like a quantum dot. We note that trapping of neutral cesium atoms in a one-dimensional optical lattice above the surface of a nanofiber has been realized experimentally \cite{twocolor experiment}. With the above assumption, we can neglect the kinetic energy and
the potential in the Hamiltonian (\ref{30}). Then, we obtain the following effective Hamiltonian:
\begin{eqnarray}\label{30c}
H_{\mathrm{eff}}&=&-\frac{1}{2}\hbar\Delta_a\sigma_z-\hbar\Delta_c a^\dagger a
-i\hbar G(a \sigma^\dagger-a^\dagger \sigma)
\nonumber\\&&\mbox{} 
-i\hbar\eta( a-a^{\dagger}).
\end{eqnarray}
We emphasize that the effect of the van der Waals potential on the atomic transition frequency $\omega_a$ and the detuning $\Delta_c$ is kept in the above Hamiltonian. It is clear that the effect of the van der Waals potential on the atomic transition frequency is negligible in the region of large atom-to-surface distances $r-a$ but significant in the region of small distances $r-a$. In our numerical calculations presented in Sec. \ref{sec:results},  we assume for simplicity that the van der Waals potential in the case of the fiber is the same as that in the case of a flat surface, that is, $V_{\alpha}=-C_{3\alpha}/(r-a)^3$, where $\alpha=g,e$.

\section{Density-matrix equations}
\label{sec:equation}

Let $\rho$ be the density operator for the combined atom-field system. In the presence of the
atomic decay and the cavity damping, the time evolution of $\rho$ is governed by the master equation
\begin{eqnarray}\label{18}
\dot{\rho}&=&\frac{i}{\hbar}[\rho,H_{\mathrm{eff}}]-\frac{\gamma}{2}(\sigma^\dagger\sigma\rho-2\sigma\rho\sigma^\dagger+\rho\sigma^\dagger\sigma)
\nonumber\\&&\mbox{}
-\frac{\kappa}{2}(a^\dagger a\rho-2a\rho a^\dagger+\rho a^\dagger a).
\end{eqnarray}
Here $\gamma=\gamma_{\mathrm{gyd}}+\gamma_{\mathrm{rad}}$ is the decay rate of the atom in the presence
of the nanofiber and in the absence of the FBG cavity. 

We use the basis $|\alpha,n\rangle$ formed from the internal states $|\alpha\rangle=|e\rangle,|g\rangle$ of the atom and the number states $|n\rangle$ of the guided cavity field.
For the matrix elements $\rho_{\alpha,n;\alpha',n'}=\langle\alpha,n |\rho|\alpha',n'\rangle$ of the density operator $\rho$, we find the equations
\begin{eqnarray}\label{36a}
\lefteqn{\dot{\rho}_{e,n;e,n'}=-i\Delta_c(n'-n)\rho_{e,n;e,n'}}
\nonumber\\&&\mbox{}
-G(\sqrt{n'+1}\rho_{e,n;g,n'+1}+\sqrt{n+1}\rho_{g,n+1;e,n'})
\nonumber\\&&\mbox{}
+\eta(\sqrt{n'}\rho_{e,n;e,n'-1}+\sqrt{n}\rho_{e,n-1;e,n'}
\nonumber\\&&\mbox{}
-\sqrt{n'+1}\rho_{e,n;e,n'+1}-\sqrt{n+1}\rho_{e,n+1;e,n'})
\nonumber\\&&\mbox{}
-\gamma\rho_{e,n;e,n'}
-\frac{\kappa}{2}\big[(n'+n)\rho_{e,n;e,n'}
\nonumber\\&&\mbox{}
-2\sqrt{(n'+1)(n+1)}\rho_{e,n+1;e,n'+1}\big],
\nonumber\\
\lefteqn{\dot{\rho}_{g,n;g,n'}=-i\Delta_c(n'-n)\rho_{g,n;g,n'}}
\nonumber\\&&\mbox{}
+G(\sqrt{n'}\rho_{g,n;e,n'-1}+\sqrt{n}\rho_{e,n-1;g,n'})
\nonumber\\&&\mbox{}
+\eta(\sqrt{n'}\rho_{g,n;g,n'-1}
+\sqrt{n}\rho_{g,n-1;g,n'}
\nonumber\\&&\mbox{}
-\sqrt{n'+1}\rho_{g,n;g,n'+1}
-\sqrt{n+1}\rho_{g,n+1;g,n'})
\nonumber\\&&\mbox{}
+\gamma\rho_{e,n;e,n'}
-\frac{\kappa}{2}\big[(n'+n)\rho_{g,n;g,n'}
\nonumber\\&&\mbox{}
-2\sqrt{(n'+1)(n+1)}\rho_{g,n+1;g,n'+1}\big],
\nonumber\\
\lefteqn{\dot{\rho}_{g,n;e,n'}=-i\Delta_a\rho_{g,n;e,n'}-i\Delta_c(n'-n)\rho_{g,n;e,n'}}
\nonumber\\&&\mbox{}
-G(\sqrt{n'+1}\rho_{g,n;g,n'+1}-\sqrt{n}\rho_{e,n-1;e,n'})
\nonumber\\&&\mbox{}
+\eta(\sqrt{n'}\rho_{g,n;e,n'-1}
+\sqrt{n}\rho_{g,n-1;e,n'}
\nonumber\\&&\mbox{}
-\sqrt{n'+1}\rho_{g,n;e,n'+1}
-\sqrt{n+1}\rho_{g,n+1;e,n'})
\nonumber\\&&\mbox{}
-\frac{\gamma}{2}\rho_{g,n;e,n'}
-\frac{\kappa}{2}\big[(n'+n)\rho_{g,n;e,n'}
\nonumber\\&&\mbox{}
-2\sqrt{(n'+1)(n+1)}\rho_{g,n+1;e,n'+1}\big].
\end{eqnarray}

The master equation (\ref{18}) is equivalent to the following equation for the mean value $\langle\mathcal{O}\rangle=\mathrm{Tr}\, (\mathcal{O}\rho)$ of an arbitrary operator $\mathcal{O}$:
\begin{eqnarray}\label{18a}
\langle\dot{\mathcal{O}}\rangle&=&\frac{i}{\hbar}\langle[H_{\mathrm{eff}},\mathcal{O}]\rangle
-\frac{\gamma}{2}\langle\mathcal{O}\sigma^\dagger\sigma-2\sigma^\dagger\mathcal{O}\sigma+\sigma^\dagger\sigma\mathcal{O}\rangle
\nonumber\\&&\mbox{}
-\frac{\kappa}{2}\langle \mathcal{O}a^\dagger a-2 a^\dagger\mathcal{O}a\rho+a^\dagger a\mathcal{O}\rangle.
\end{eqnarray}
In particular, we find the equations
\begin{eqnarray}\label{37}
\frac{d}{dt}\langle a\rangle&=&i\Delta_c\langle a\rangle+G\langle\sigma\rangle
-\frac{\kappa}{2}\langle a\rangle+\eta,
\nonumber\\
\frac{d}{dt}\langle\sigma\rangle&=&i\Delta_a\langle \sigma\rangle+G\langle a\sigma_z\rangle-\frac{\gamma}{2}\langle \sigma\rangle,
\end{eqnarray}
and
\begin{eqnarray}\label{38}
\lefteqn{\frac{d}{dt}\langle a^\dagger a\rangle=G\langle a^\dagger\sigma+a\sigma^\dagger\rangle
-\kappa\langle a^\dagger a\rangle
+\eta(\langle a^\dagger\rangle+\langle a\rangle),}
\nonumber\\
\lefteqn{\frac{d}{dt}\langle \sigma^\dagger\sigma\rangle=-G\langle a^\dagger\sigma+a\sigma^\dagger\rangle-\gamma\langle \sigma^\dagger\sigma\rangle,}
\nonumber\\
\lefteqn{\frac{d}{dt}\langle a^\dagger \sigma+a\sigma^\dagger\rangle=
i\Delta\langle a^\dagger \sigma-a\sigma^\dagger\rangle}
\nonumber\\&&\mbox{}
+2G\langle a^\dagger a \sigma_z+\sigma^\dagger\sigma\rangle
-\frac{\kappa+\gamma}{2}\langle a^\dagger \sigma+a\sigma^\dagger\rangle
\nonumber\\&&\mbox{}
+\eta(\langle\sigma\rangle+\langle\sigma^\dagger\rangle),
\nonumber\\
\lefteqn{\frac{d}{dt}\langle a^\dagger \sigma-a\sigma^\dagger\rangle=
i\Delta\langle a^\dagger \sigma+a\sigma^\dagger\rangle}
\nonumber\\&&\mbox{}
-\frac{\kappa+\gamma}{2}\langle a^\dagger \sigma-a\sigma^\dagger\rangle
+\eta(\langle\sigma\rangle-\langle\sigma^\dagger\rangle).
\end{eqnarray}
Here $\Delta=\omega_c-\omega_a$ is the cavity--atom detuning.

In order to get insight into the effect of the atom on the guided cavity field, we use the procedures of Ref. \cite{Ritsch} to linearize Eqs. (\ref{37}) and (\ref{38}). For this purpose, we assume that the probe field is so weak that the excited state is hardly occupied and there is at most one photon in the cavity. In this case, we have \cite{Ritsch}
\begin{eqnarray}\label{39}
\langle a\sigma_z \rangle=-\langle a\rangle,
\nonumber\\
\langle a^\dagger a\sigma_z \rangle=-\langle a^\dagger a\rangle.
\end{eqnarray}
With the help of the above formulae, we can linearize Eqs. (\ref{37}) and (\ref{38}) and solve them
in the steady-state regime. The results are \cite{Ritsch}
\begin{eqnarray}\label{42}
&&\langle a\rangle=-\frac{\eta}{D}(i\Delta_a-\gamma/2),
\nonumber\\
&&\langle \sigma\rangle=-\frac{\eta}{D}G,
\nonumber\\
&&\langle a^\dagger a\rangle=\frac{\eta^2}{|D|^2}(\Delta_a^2+\gamma^2/4),
\nonumber\\
&&\langle \sigma^\dagger \sigma\rangle=\frac{\eta^2}{|D|^2}G^2,
\nonumber\\
&&\langle a^\dagger \sigma+a\sigma^\dagger\rangle=-\frac{\eta^2}{|D|^2}G\gamma,
\nonumber\\
&&\langle a^\dagger \sigma-a\sigma^\dagger\rangle=-2i\frac{\eta^2}{|D|^2}G\Delta_a,
\end{eqnarray}
where
\begin{equation}\label{43}
D=G^2+\kappa\gamma/4-\Delta_c\Delta_a-i(\Delta_c\gamma+\Delta_a\kappa)/2.
\end{equation}
It is interesting to note that $\langle a^\dagger a\rangle=\langle a^\dagger\rangle\langle a\rangle$,
$\langle \sigma^\dagger \sigma\rangle=\langle \sigma^\dagger\rangle\langle\sigma\rangle$, and
$\langle a^\dagger \sigma\rangle=\langle a^\dagger\rangle\langle\sigma\rangle$. These relations are valid only in the case of a weakly driven cavity. The explicit expressions for the mean photon number $N_{\mathrm{cav}}=\langle a^{\dagger}a\rangle$ and the atomic excited-state population $P_e=\langle \sigma^{\dagger}\sigma\rangle$ can be represented in the forms
\begin{equation}\label{44a}
N_{\mathrm{cav}}=\frac{\eta^2(\Delta_a^2+\gamma^2/4)}
{(G^2+\kappa\gamma/4-\Delta_c\Delta_a)^2+(\Delta_c\gamma+\Delta_a\kappa)^2/4}
\end{equation}
and
\begin{equation}\label{44b}
P_e=\frac{\eta^2G^2}{(G^2+\kappa\gamma/4-\Delta_c\Delta_a)^2+(\Delta_c\gamma+\Delta_a\kappa)^2/4},
\end{equation}
respectively. Since the parameters $G$, $\gamma$, and $\Delta_a$ vary in space, the intensity of the transmitted light depends on the position of the atom. We note from Eq. (\ref{44a}) that the atom can significantly affect the guided cavity field if the condition $G^2\gg \max(\kappa\gamma/4,|\Delta_c||\Delta_a|,|\Delta_c|\gamma/2,|\Delta_a|\kappa/2)$ is satisfied. It has been shown that, in the case of a high-finesse microcavity driven by a weak probe field, under the condition of strong coupling, the position of the atom can be inferred from the intensity of the transmitted light \cite{Rempe,Doherty,Mabuchi,Hood,Munstermann}.

\section{Numerical results}
\label{sec:results}

In this section, we present our numerical results. We use a truncated basis to solve
numerically the density-matrix equations (\ref{36a}) in the steady-state regime. From the steady-state solution for the density matrix $\rho$, we calculate the mean number $N_{\mathrm{cav}}=\langle a^{\dagger}a\rangle$ of photons in the guided cavity field and the population $P_e=\langle \sigma^{\dagger}\sigma\rangle$ of the atomic excited state. We also calculate the second-order photon correlation function $g_{\mathrm{cav}}^{(2)}=\langle a^{\dagger}a^{\dagger}aa\rangle/\langle a^{\dagger}a\rangle^2$, which characterizes the photon statistics \cite{Quantum Optics}. It is known that the photon distribution of a coherent state is a Poisson distribution, which gives $g_{\mathrm{cav}}^{(2)}=1$. When $g_{\mathrm{cav}}^{(2)}>1$ or $g_{\mathrm{cav}}^{(2)}<1$, the photon statistics is said to be super- or sub-Poissonian, respectively. The occurrence of sub-Poissonian photon statistics, indicated by the inequality $g_{\mathrm{cav}}^{(2)}<1$, means that the state of the field is nonclassical \cite{Quantum Optics}.

\begin{figure}[tbh]
\begin{center}
 \includegraphics{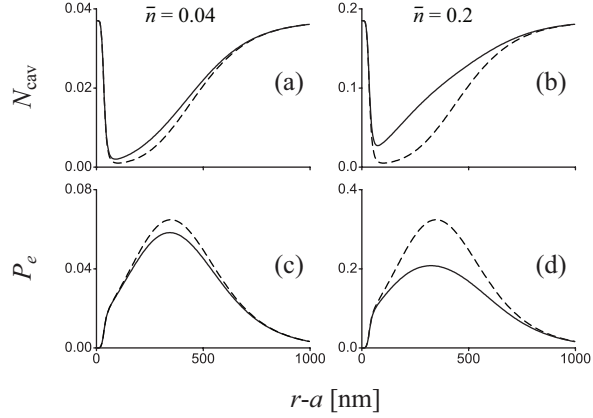}
 \end{center}
\caption{Mean photon number $N_{\mathrm{cav}}$ and atomic excited-state population $P_e$ as functions of the atom-to-surface distance $r-a$. The solid and dashed lines represent the results of the calculations from the exact steady-state solutions to the density-matrix equations (\ref{36a}) and from the approximate solutions (\ref{44a}) and (\ref{44b}), respectively. The fiber radius is $a=200$ nm. The FBG cavity length is $L=10$ cm. The FBG mirror reflectivity is $|R|^2=0.9$. The input probe power is $P_{\mathrm{in}}=1$ pW (left panel) and 5 pW (right panel). The parameters for the two-level atom correspond to the $D_2$-line transition $6S_{1/2}F=4\,M=4\leftrightarrow 6P_{3/2}F'=5\,M'=5$ of atomic cesium, with the wavelength $\lambda_0=852$ nm and the natural linewidth $\gamma_0=5.25$ MHz. The frequencies of the probe field, the cavity, and the atom in free space are equal, i.e. $\omega_p=\omega_c=\omega_0$. The order $m$ of the cavity resonance mode is an even number. The axial position $z$ of the atom corresponds to an antinode of the cavity standing-wave field. The van der Waals coefficients are
$C_{3g}=1.56$ kHz $\mu$m$^3$ and $C_{3e}=3.09$ kHz $\mu$m$^3$.}
\label{fig2}
\end{figure}

We plot in Fig. \ref{fig2} the mean number $N_{\mathrm{cav}}$ of intracavity photons and the population $P_e$ of the atomic excited state as functions of the distance $r-a$ from the atom to the fiber surface. The FBG mirror reflectivity is $|R|^2=0.9$, which corresponds to the finesse $F=\pi |R|/(1-|R|^2)\cong 30$, a moderate value. The FBG cavity length is $L=10$ cm. The frequencies of the probe field, the cavity, and the atom in free space are equal, i.e. $\omega_p=\omega_c=\omega_0$. The input probe power is $P_{\mathrm{in}}=1$ pW (left panel) and 5 (right panel) pW, corresponding to the mean number $\bar{n}=0.04$ and 0.2, respectively, of intracavity photons in the absence of the atom. The order $m$ of the cavity resonance mode is an even number. For the parameters of this figure, we find the atom-cavity coupling coefficient $G(r,z)|_{r=a,z=0}=5.33\gamma_0$, the cavity damping rate $\kappa=7.02\gamma_0$, and the atomic decay rate $\gamma(r)|_{r=a}=1.73\gamma_0$, with $\gamma_0=5.25$ MHz being the natural linewidth of the $D_2$-line transition of atomic cesium. It is clear that the strong-coupling condition $2|G|>\kappa,\gamma$ is satisfied when the atom is close to the fiber and to an antinode of the cavity standing-wave field. The figure shows that $N_{\mathrm{cav}}$ and $P_e$ substantially depend on the atom-to-surface distance $r-a$. We observe that the radial profile of $N_{\mathrm{cav}}$ has a minimum and the radial profile of $P_e$ has a maximum. The minimum of $N_{\mathrm{cav}}$ is closer to the surface than the maximum of $P_e$ is. When $r-a$ is large, $N_{\mathrm{cav}}$ decreases and $P_e$ increases with decreasing $r-a$. Such behaviors are due to the radial dependence of the atom-cavity coupling coefficient $G$. In contrast, when $r-a$ is small, $N_{\mathrm{cav}}$ increases and $P_e$ decreases with decreasing $r-a$. Such behaviors are due to the effect of the van der Waals potential on the radial dependence of the surface-shifted atomic transition frequency $\omega_a$. We emphasize again that the van der Waals potential is negligible in the region of large $r-a$ but significant in the region of small $r-a$. Figure \ref{fig2}(d) shows that, although the probe field is very weak ($P_{\mathrm{in}}=5$ pW) and the mean photon number in the absence of the atom is small ($\bar{n}=0.2$), the atomic excitation $P_e$ can reach quite large values (e.g. $P_e\cong0.2$ at $r-a\cong320$ nm). Such substantial excitation is a result of the strong coupling between the atom and the guided cavity field. Comparison between the solid and dashed lines in Fig. \ref{fig2} shows that the approximate solutions (\ref{44a}) and (\ref{44b}) almost coincide with the results of the exact calculations from the density-matrix equations (\ref{36a}) when $\bar{n}$ is on the order of or smaller than 0.04, that is, when the power $P_{\mathrm{in}}$ of the probe field is on the order of or smaller than 1 pW.  When the power $P_{\mathrm{in}}$ of the probe field is on the order of or larger than 5 pW, the discrepancy between the exact and approximate solutions becomes serious.

\begin{figure}[tbh]
\begin{center}
 \includegraphics{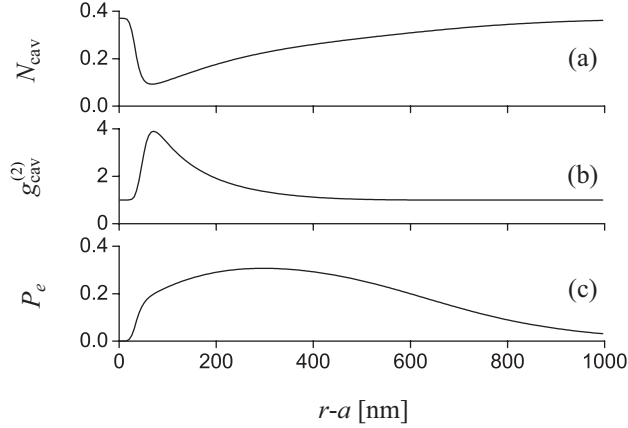}
 \end{center}
\caption{Mean photon number $N_{\mathrm{cav}}$, second-order photon correlation function $g_{\mathrm{cav}}^{(2)}$, and atomic excited-state population $P_e$ as functions of the atom-to-surface distance $r-a$. The input probe power is $P_{\mathrm{in}}=10$ pW. The FBG cavity length is $L=10$ cm. The frequencies of the probe field, the cavity, and the atom in free space are equal, i.e. $\omega_p=\omega_c=\omega_0$. The axial position $z$ of the atom corresponds to an antinode of the cavity standing-wave field. Other parameters are as in Fig.~\ref{fig2}.}
\label{fig3}
\end{figure}

The effect of the atom on the quantum guided cavity field is significant only when the mean photon number is small. However, it is not easy to measure the transmitted field when the mean intracavity photon number is too small. Therefore, we focus on the case where the mean photon number in the absence of the atom is less than one but not too small. For this purpose, we choose 
the input probe power $P_{\mathrm{in}}=10$ pW, which corresponds to the value $\bar{n}=0.4$ for the mean photon number in the absence of the atom. The FBG cavity length is $L=10$ cm.
We plot in Fig. \ref{fig3} the mean photon number $N_{\mathrm{cav}}$, the second-order photon correlation function $g_{\mathrm{cav}}^{(2)}$, and the atomic excited-state population $P_e$ as functions of the distance $r-a$ from the atom to the fiber surface. The frequencies of the probe field, the cavity, and the atom in free space are equal, i.e. $\omega_p=\omega_c=\omega_0$. Figures \ref{fig3}(a) and \ref{fig3}(c) show that $N_{\mathrm{cav}}$ and $P_e$ substantially depend on the atom-to-surface distance $r-a$. Figure \ref{fig3}(b) shows that $g_{\mathrm{cav}}^{(2)}> 1$, that is, the photon statistics is super-Poissonian. It is clear from the figure that, in the region of large $r-a$, where the van der Waals potential is weak, the second-order photon correlation function $g_{\mathrm{cav}}^{(2)}$ increases with decreasing $r-a$. However, in the region of small $r-a$, where the van der Waals potential is significant, $g_{\mathrm{cav}}^{(2)}$ decreases with decreasing $r-a$. Thus, the $r$ dependence of $g_{\mathrm{cav}}^{(2)}$ is basically opposite to that of $N_{\mathrm{cav}}$. Note that in the vicinity around the distance $r-a=70$ nm, the correlation function $g_{\mathrm{cav}}^{(2)}$ can become larger than 2, that is, the photon distribution of the guided cavity field can become broader than the Boltzmann distribution for thermal states.

\begin{figure}[tbh]
\begin{center}
 \includegraphics{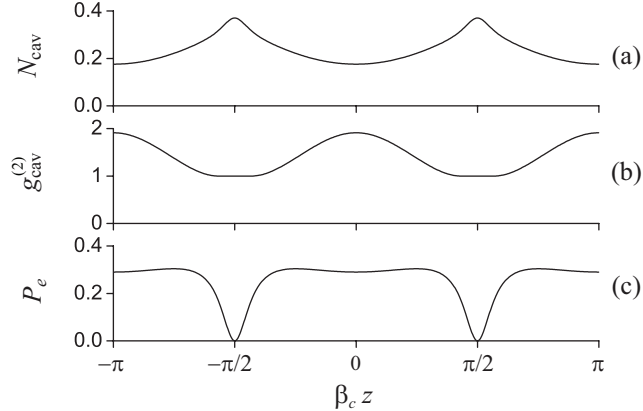}
 \end{center}
\caption{Mean photon number $N_{\mathrm{cav}}$, second-order photon correlation function $g_{\mathrm{cav}}^{(2)}$, and atomic excited-state population $P_e$ as functions of the axial position $z$ of the atom along the fiber. The distance from the atom to the fiber surface is $r-a=200$ nm. Other parameters are as for Fig. \ref{fig3}.}
\label{fig4}
\end{figure}

We plot in Fig. \ref{fig4} the mean photon number $N_{\mathrm{cav}}$, the second-order photon correlation function $g_{\mathrm{cav}}^{(2)}$, and the atomic excited-state population $P_e$ as functions of the axial position $z$ of the atom along the fiber. Figure \ref{fig4}(a) shows that $N_{\mathrm{cav}}$ has minima and maxima at the antinodes and the nodes of the cavity standing-wave field, respectively. It is clear that, in the fiber axial direction, $N_{\mathrm{cav}}$ follows the spatial  oscillations of the coupling coefficient $G$ [see Eq. (\ref{44a})]. Figure \ref{fig4}(b) shows that $g_{\mathrm{cav}}^{(2)}$ has maxima at the antinodes of the guided cavity field. The figure also shows that $g_{\mathrm{cav}}^{(2)}>1$, that is, the photon statistics is super-Poissonian, in a broad region around each antinode, and that $g_{\mathrm{cav}}^{(2)}\cong1$, that is, the photon statistics is almost Poissonian, in a smaller region around each node. A careful look at the data reveals that $g_{\mathrm{cav}}^{(2)}<1$, that is, the photon statistics is sub-Poissonian, when the position of the atom is slightly deviated from a node. The appearance of such a nonclassical state is related to the fact that
the absorption of a photon of a weak field by a ground-state atom (or a weakly excited one) can reduce the photon-number spread. Figure \ref{fig4}(c) shows that the axial position dependence of $P_e$ is complicated [see Eq. (\ref{44b})]. Indeed, $P_e$ has shallow minima at the antinodes and deep minima at the nodes of the guided cavity field.

\begin{figure}[tbh]
\begin{center}
 \includegraphics{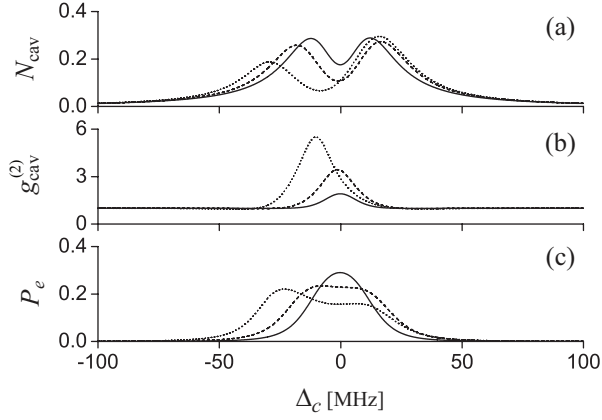}
 \end{center}
\caption{Mean photon number $N_{\mathrm{cav}}$, second-order photon correlation function $g_{\mathrm{cav}}^{(2)}$, and atomic excited-state population $P_e$ as functions of the detuning $\Delta_c$ of the probe field from the cavity resonance. The cavity is at exact resonance with the atom in free space, i.e. $\omega_c=\omega_0$. The distance from the atom to the fiber surface is $r-a=200$ nm (solid lines), 100 nm (dashed lines), and 50 nm (dotted lines). The axial position $z$ of the atom along the fiber corresponds to an antinode of the cavity standing-wave field. Other parameters are as for Fig. \ref{fig3}.}
\label{fig5}
\end{figure}

We plot in Fig. \ref{fig5} the mean photon number $N_{\mathrm{cav}}$, the second-order photon correlation function $g_{\mathrm{cav}}^{(2)}$, and the atomic excited-state population $P_e$ as functions of the detuning $\Delta_c$ of the probe field from the cavity resonance in the case where the cavity is at exact resonance with the atom in free space, i.e. $\omega_c=\omega_0$. Figure \ref{fig5}(a) shows that the spectrum of $N_{\mathrm{cav}}$ has a vacuum Rabi splitting. The positions of the two peaks are close to $\Delta_{\mathrm{vdW}}/2\pm \sqrt{G^2+\kappa\gamma/4+\Delta_{\mathrm{vdW}}^2/4}$, where $\Delta_{\mathrm{vdW}}=V_{eg}/\hbar$ is the frequency shift caused by the van der Waals potential. Due to the effect of the van der Waals potential, the two peaks are not symmetric in height and position. The peak on the negative side of the detuning $\Delta_c$ is lower and farther away from the center than the other peak. Comparison between the solid lines (for $r-a=200$ nm), the dashed lines
(for $r-a=100$ nm), and the dotted lines  (for $r-a=50$ nm) shows that the smaller the distance $r-a$, the stronger the asymmetry of the peaks. The reason is that the depth of the van der Waals potential increases with decreasing distance $r-a$. Figure \ref{fig5}(b) shows that $g_{\mathrm{cav}}^{(2)}>1$, that is, the photon statistics is super-Poissonian, in a broad frequency region where the corresponding atomic excitation $P_e$ is significant [see Fig. \ref{fig5}(c)]. The dashed and dotted lines in Figs. \ref{fig5}(b) indicate that the correlation function  $g_{\mathrm{cav}}^{(2)}$ can become larger than 2, that is, the photon distribution of the guided cavity field can become broader than the Boltzmann distribution for thermal states. Such a broadening of the photon distribution is a consequence of the emission from the excited atom.    

\begin{figure}[tbh]
\begin{center}
 \includegraphics{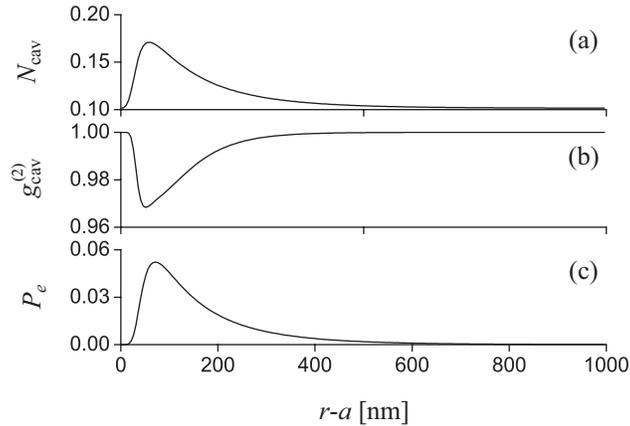}
 \end{center}
\caption{Same as Fig. \ref{fig3} except for the detuning $\Delta_c=30$ MHz.}
\label{fig6}
\end{figure}

\begin{figure}[tbh]
\begin{center}
 \includegraphics{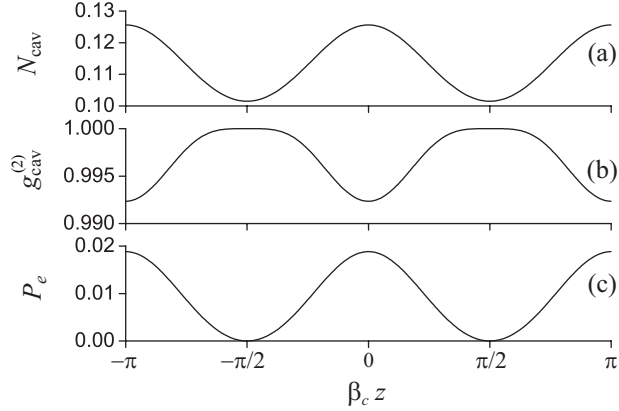}
 \end{center}
\caption{Same as Fig. \ref{fig4} except for the detuning $\Delta_c=30$ MHz.}
\label{fig7}
\end{figure}

According to Fig. \ref{fig5}, the effect of the atom on the guided cavity field depends on the detuning of the probe field. To see the contrast between the case of exact resonance and the case of  substantial detuning, we plot in Figs. \ref{fig6} and \ref{fig7} the results of the numerical calculations for the case where $\Delta_c=30$ MHz. Figure \ref{fig6}(a) shows that, when $r-a$ is large, $N_{\mathrm{cav}}$ increases with decreasing $r-a$. In this region, the presence of the atom increases the number of intracavity photons. Figure \ref{fig7}(a) shows that the number of intracavity photons is largest when the atom is positioned at an antinode of the guided cavity field. Figures \ref{fig6}(b) and \ref{fig7}(b) show that, outside the nodes of the cavity standing-wave field, we have $g_{\mathrm{cav}}^{(2)}<1$, that is, the photon statistics is sub-Poissonian. Comparison between Figs. \ref{fig3} and \ref{fig6} and between Figs. \ref{fig4} and \ref{fig7} shows that the two cases have very different behaviors.

\begin{figure}[tbh]
\begin{center}
 \includegraphics{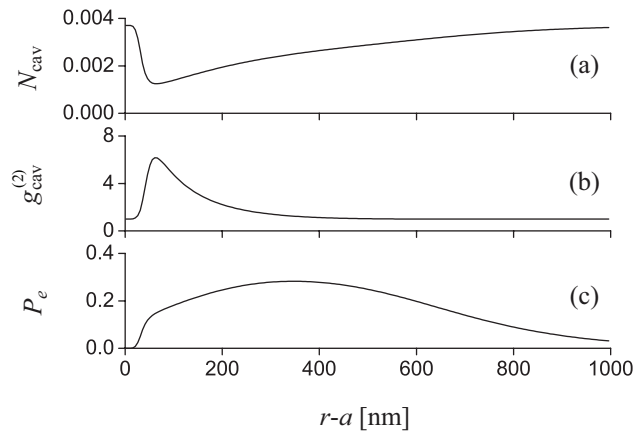}
 \end{center}
\caption{Same as Fig. \ref{fig3} except for the cavity length $L=1$ mm.}
\label{fig8}
\end{figure}
\begin{figure}[tbh]
\begin{center}
 \includegraphics{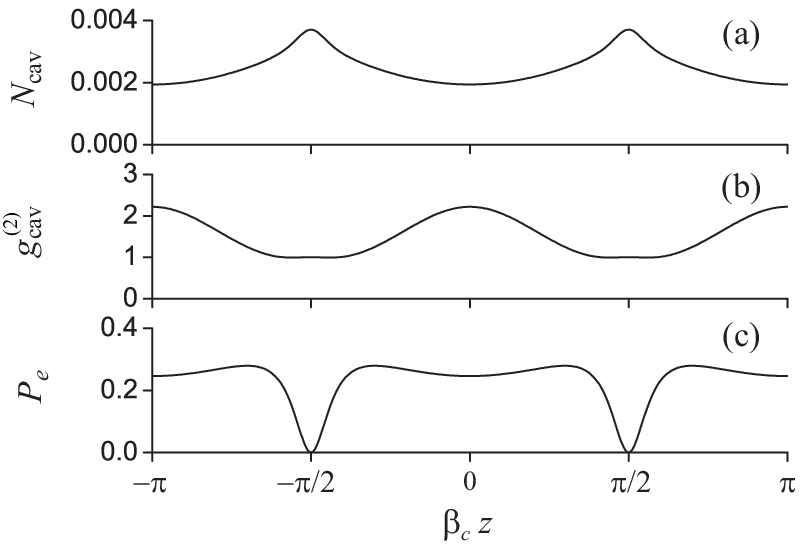}
 \end{center}
\caption{Same as Fig. \ref{fig4} except for the cavity length $L=1$ mm.}
\label{fig9}
\end{figure}
\begin{figure}[tbh]
\begin{center}
 \includegraphics{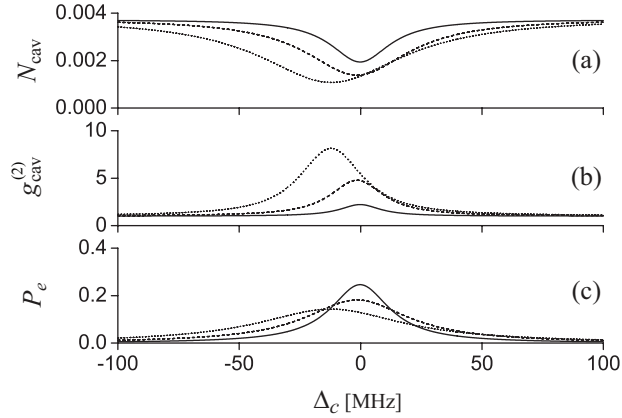}
 \end{center}
\caption{Same as Fig. \ref{fig5} except for the cavity length $L=1$ mm.}
\label{fig10}
\end{figure}

According to Eq. (\ref{44a}), the atom can significantly affect the guided cavity field if the condition $G^2\gg \max(\kappa\gamma/4,|\Delta_c||\Delta_a|,|\Delta_c|\gamma/2,|\Delta_a|\kappa/2)$ is satisfied. In the case where $\Delta_c,\Delta_a\cong0$, the above condition reduces to $G^2\gg \kappa\gamma/4$. The strong-coupling condition $2G\gg\kappa,\gamma$ is not required. We plot in Figs. \ref{fig8}--\ref{fig10} the results of the numerical calculations for the case where the FBG cavity length is $L=1$ mm. Such a length is two orders smaller than the length used for the calculations of the previous figures. For the parameters of Figs. \ref{fig8}--\ref{fig10}, we find the atom-cavity coupling coefficient $G(r,z)|_{r=a,z=0}=53.3\gamma_0$, the cavity damping rate $\kappa=702\gamma_0$, and the atomic decay rate $\gamma(r)|_{r=a}=1.73\gamma_0$. Since the cavity length $L$ is small, the atom-cavity coupling coefficient $G$ and the cavity damping rate $\kappa$ are large. 
It is clear that the strong-coupling condition $2G\gg\kappa,\gamma$ is not satisfied. 
The reason is that $\kappa$ increases with decreasing $L$ faster than $G$ does. Figures \ref{fig8}--\ref{fig10} show that the atom can still affect significantly the guided cavity field even in the overdamped regime. Note that the shapes of the curves in Figs. \ref{fig8} and \ref{fig9} are very similar to those in Figs. \ref{fig3} and \ref{fig4}, respectively. The substantial difference between the magnitudes of the mean photon number $N_{\mathrm{cav}}$ in the case of Figs. \ref{fig8}--\ref{fig10} and the case of Figs. \ref{fig3}--\ref{fig5} results from the substantial difference in the cavity length $L$. Meanwhile, the shapes of the curves in Fig. \ref{fig10}(a) are very different from the shapes of the curves in Fig. \ref{fig5}(a). Indeed, the curves in Fig. \ref{fig10}(a) do not show the vacuum Rabi splitting.  Such a splitting can be observed only under the condition of strong coupling.

\section{SUMMARY}
\label{sec:summary}

We have studied the effect of an atom on a quantum guided field in a weakly driven FBG cavity. We have calculated the mean photon number, the second-order photon correlation function, and the atomic excited-state population. We have shown that, due to the confinement of the guided cavity field in the fiber cross-section plane and in the space between the FBG mirrors, the presence of the atom in the FBG cavity can significantly affect the mean photon number as well as the photon statistics 
even though the cavity finesse is moderate, the cavity is long, and the probe field is weak.
Due to the effect of the van der Waals potential, the vacuum Rabi splitting can become asymmetric with respect to the positions and heights of the peaks.
The photon statistics of the quantum guided cavity field can be super- or sub-Poissonian depending on the position of the atom and the detuning of the probe field.

\end{document}